\documentclass{article}
\usepackage{spconf,amsmath,graphicx}
\usepackage{multicol}
\usepackage{amsmath,amssymb,cite,multirow,subfigure}
\usepackage{graphicx}
\usepackage{url}
\usepackage[linesnumbered,ruled, vlined]{algorithm2e}
\usepackage{algorithmic}
\usepackage{makecell}

\usepackage{booktabs}
\usepackage{threeparttable}

\title{A HYBRID PARTIAL SUM COMPUTATION UNIT ARCHITECTURE FOR LIST DECODERS OF POLAR CODES}
\name{Jun Lin and Zhiyuan Yan}
\address{Department of Electrical and Computer Engineering, Lehigh University, PA, USA}
\begin{document}
%
\maketitle
\begin{abstract}
Although the successive cancelation (SC) algorithm works well for very long polar codes, its error performance for shorter polar codes is much worse. Several SC based list decoding algorithms have been proposed to improve the error performances of both long and short polar codes. A significant step of SC based list decoding algorithms is the updating of partial sums for all decoding paths.
In this paper, we first proposed a lazy copy partial sum computation algorithm for SC based list decoding algorithms. Instead of copying partial sums directly, our lazy copy algorithm copies indices of partial sums. Based on our lazy copy algorithm, we propose a hybrid partial sum computation unit architecture, which employs both registers and memories so that the overall area efficiency is improved. Compared with a recent partial sum computation unit for list decoders, when the list size $L=4$, our partial sum computation unit achieves an area saving of 23\% and 63\% for block length $2^{13}$ and $2^{15}$, respectively.
\end{abstract}
\begin{keywords}
Polar codes, list decoding, partial sum computation
\end{keywords}
\section{Introduction}
\label{sec:intro}

Polar codes~\cite{arikan} are a significant breakthrough in coding theory, since they can provably achieve channel capacity.
Several successive cancelation (SC) based list decoding algorithms have been proposed to improve the error performances of both long and short polar codes.
An SC list (SCL) decoding algorithm, recently proposed in~\cite{ido_list1}, performs better than the SC algorithm. While the SCL algorithm in~\cite{ido_list1} selects the output codeword from $L$ candidates, where $L$ is the list size, based on path metric only, this selection is aided by using the cyclic redundancy check (CRC) in ~\cite{ido_list2,list2, list3}.
A CRC-aided SCL (CA-SCL) algorithm performs much better than the SCL algorithm at the expense of negligible loss in code rate. A log-likelihood ratio (LLR) based SCL decoding algorithm was proposed in~\cite{llr_list} to reduce the message memory area of a SCL or CA-SCL decoder. In~\cite{jun_sips}, we proposed an LLR based list decoding algorithm with reduced latency for polar codes. In~\cite{gabi_low_latency}, an increased speed polar list decoder was also proposed.

Inspired by their superior error performances, the SCL and CA-SCL list decoder architectures for polar codes were discussed in~\cite{tree_list_dec,jun_list,jun_low_mem_list}, where the partial sum computation units were based on registers. When the corresponding block length is large (e.g. $N=2^{15}$), the main drawbacks of the register based partial sum computation architectures are the area overhead and the power dissipation due to the copying of partial sums.

In this paper, we first propose a lazy copy partial sum computation algorithm, which copies only path indices instead of partial sums.
We also propose a hybrid partial sum computation architecture for list decoders of polar codes. Our architecture employs static RAMs (SRAMs) or register files (RFs) to reduce the area overhead when $N$ is large.
Compared with the partial sum architecture shown in~\cite{jun_low_mem_list}, when the list size $L=4$, our partial sum computation unit achieves an area saving of 23\% and 63\% for block length $2^{13}$ and $2^{15}$, respectively. It seems that our partial sum computation unit architecture is more suitable for large block length.

The proposed partial sum computation unit architecture works for all SC based list decoding algorithms mentioned above. Compared with the partial sum computation in the SCL and CA-SCL decoding algorithms~\cite{ido_list1, ido_list2}, the input to the partial sum computation of the list decoding algorithm in~\cite{jun_sips, gabi_low_latency} may be a bit vector instead of a single bit.
The lazy copy scheme proposed here is different from that proposed in~\cite{ido_list2}, which needs complex array index computation and is not hardware efficient.
Our partial sum computation unit is based on lazy copy, and is different from those in~\cite{tree_list_dec,jun_low_mem_list, chuan_list, yuan_low_latency}, which are based on direct copy.
Besides, the partial sum computation unit architecture was not investigated in~\cite{jun_sips, gabi_low_latency}.

The rest of the paper is organized as follows. In Section~\ref{sec: background}, some background information is reviewed. The proposed hybrid partial sum computation unit architecture is discussed in Section~\ref{sec: top_archi}. The implementation results are shown in Section~\ref{sec: hardware_imp}. At last, the conclusions are drawn in Section~\ref{sec: conclusion}.

\section{Background}\label{sec: background}
A generation matrix of a polar code is an $N\times N$ matrix $G=B_NF^{\otimes n}$, where $N=2^n$, $B_N$ is the bit reversal permutation matrix~\cite{arikan}, and $F=\left[{1\atop 1}{0\atop 1}\right]$. Here $\otimes n$ denotes the $n$th Kronecker power and $F^{\otimes n} = F\otimes F^{\otimes (n-1)}$. Let $u_0^{N-1} = (u_0,u_1,\cdots,u_{N-1})$ denote the data bit sequence and $x_0^{N-1} = (x_0,x_1,\cdots,x_{N-1})$ the corresponding encoded bit sequence, then $x_0^{N-1}=u_0^{N-1}G$.

For $l = 0,1,\cdots, L-1$ and $t=1, 2, \cdots, n$, let $C_{l,t}$ be a bit matrix of $2^{n-t}\times 2$ elements: $C_{l,t}[j][0]$ and $C_{l,t}[j][1]$ store a single bit partial sum, respectively, for $j=0,1,\cdots,2^{n-t}-1$. The partial sums corresponding to decoding path $l$ are $C_{l,n}$, $C_{l,n-1}$, $\cdots$, $C_{l,1}$~\cite{ido_list1}.

For the list decoder architectures in~\cite{tree_list_dec, jun_list, jun_low_mem_list}, all partial sums are stored in registers and the partial sums of decoding path $l'$ are copied to decoding path $l$ when decoding path $l'$ needs to be copied to decoding path $l$. More specifically, $C_{l',t}$ is copied to $C_{l,t}$ for $t=1, 2, \cdots, n$. The partial sum computation unit (PSCU) in~\cite{tree_list_dec} and~\cite{jun_list} needs $L(N-1)$ and $L(\frac{N}{2}-1)$ single bit registers to store partial sums, where $N$ is the code length and $L$ the list size. Thus, for large $N$, the register based PSCU architectures~\cite{tree_list_dec, jun_list} are inefficient for two reasons. First, the area of the PSCU is linearly proportional to $N$. For large $N$, the area of PSCU is high since registers are usually area demanding. Second, the power dissipation due to the copying of partial sums between different decoding paths is high when $N$ is large.

\section{Proposed hybrid partial sum computation unit} \label{sec: top_archi}

\subsection{Lazy copy partial sum computation}

In order to simplify the copy operations, a lazy copy partial sum computation (LCPC) algorithm is proposed in Algorithm~\ref{algo: psum_comp}, where $p_l[t]\mbox{ }(l=0,1,\cdots,L-1\mbox{ and }t=0,1,\cdots,n)$ is a list index reference. $v$ denotes a node from the decoding tree~\cite{low_latency_polar, jun_sips} of a polar code. $t_v$ denotes the layer index~\cite{jun_sips} of the node $v$. IDX$_1$ denotes the index of the last leaf node of node $v$. Let $(B_{n-1}, B_{n-2},\cdots, B_0)$ denote the binary representation of IDX$_1$, where $B_{n-1}$ is the most significant bit. $I_e = n-(j+1)$, where $j$ is an integer such that $B_r =0$ for $r \leq j$. If $B_0 \neq 0$, $I_e=n$.

\begin{algorithm}
\DontPrintSemicolon
\label{algo: psum_comp}
\SetKwInOut{Input}{input}\SetKwInOut{Output}{output}

\Input{$I_e, t_v$}
\BlankLine

\For{$t = t_v-1$ \KwTo $I_e$} {
\For{$k=0$ \KwTo $2^{n-t-1}$} {
\If{$t == I_e$}{
$C_{l,t}[2k][0] = C_{p_l[t+1],t+1}[k][0]\oplus C_{l,t+1}[k][1]$\;
$C_{l,t}[2k+1][0] = C_{l,t+1}[k][1]$\;
$p_l[t+1] = p_l[t]=l$\;
}\Else{
$C_{l,t}[2k][1] = C_{p_l[t+1],t+1}[k][0]\oplus C_{l,t+1}[k][1]$\;
$C_{l,t}[2k+1][1] = C_{l,t+1}[k][1]$\;
$p_l[t+1] = l$\;
}
}
}
\caption{LCPC Algorithm}
\end{algorithm}

In order to support the reduced latency list decoding algorithm in~\cite{jun_sips}, for a round of partial sum computation, the input is a constituent codeword~\cite{low_latency_polar, jun_sips} instead of a single binary bit~\cite{tree_list_dec, ido_list1}. Suppose a constituent codeword, $\mathcal{C}_{v,l}$, sent from or received by node $v$ for decoding path $l$ is computed, then the corresponding partial sum computations are needed. $\mathcal{C}_{v,l}$ has $2^{n-t_v}$ bits. When path $l'$ needs to be copied to path $l$, the index references are first copied before the partial sum computation shown in Alg.~\ref{algo: psum_comp} is performed. For $t=t_v, t_v-1, \cdots,0$, $p_{l'}[t]$ is copied to $p_l[t]$. If a node $v$ receives a constituent code, $p_l[t_v] = l$.

If a node $v$ sends a constituent code $\mathcal{C}_{v,l}$, it is stored in $(C_{l,t_v}[0][0]$, $C_{l,t_v}[1][0]$, $\cdots$, $C_{l,t_v}[2^{n-t_v}][0])$, and no further partial sum computations are needed.
If a node $v$ receives a constituent code $\mathcal{C}_{v,l}$, it is first stored in $(C_{l,t_v}[0][1]$, $C_{l,t_v}[1][1]$, $\cdots$, $C_{l,t_v}[2^{n-t_v}][1])$, and the remaining partial sum computations are performed with the proposed LCPC algorithm in Alg.~\ref{algo: psum_comp}, where $I_e\leq t_v-1$.

\subsection{Proposed partial sum computation unit architecture}

\begin{figure*} [hbt]
\centering
  \includegraphics[width=5.5in]{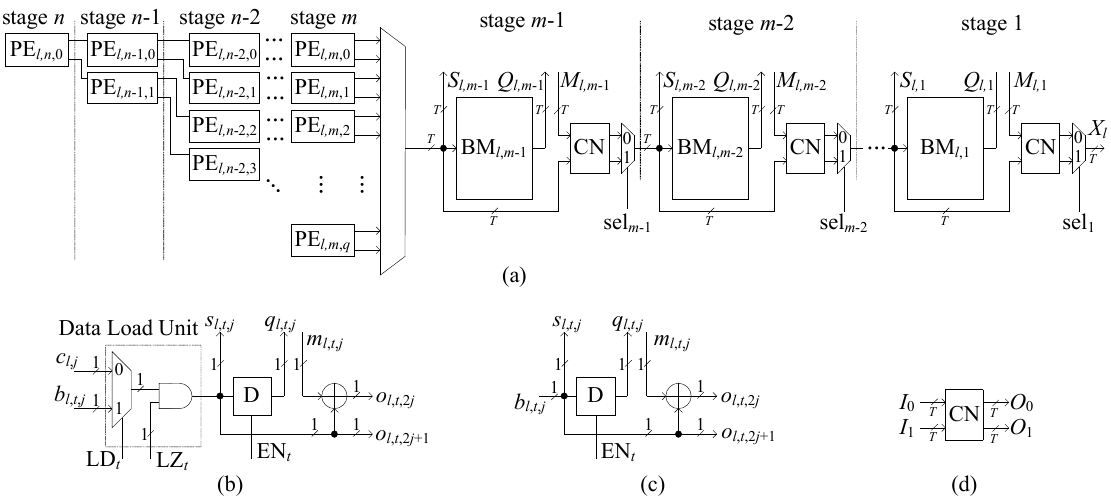}
  \caption{(a) Top architecture of the proposed PSCU. (b) Type-I PE. (c) Type-II PE. (d) Inputs and outputs of the CN.}\label{fig: psu_top}
\end{figure*}

In order to overcome the area and power overhead when $N$ is large, a hybrid partial sum computation unit (HPSU) architecture is proposed based on two improvements: (a) part of partial sums are stored in memories, while others are stored in registers, (b) the copying of partial sums is avoided by only copying list index matrices. The proposed HPSU consists of $L$ partial sum computation units. The top architecture of the proposed PSCU for decoding path $l$, shown in Fig.~\ref{fig: psu_top}(a), is described as follows.

(a) For block length $N=2^n$, the proposed PSCU consists of $n$ stages, where the first $n-m+1$ stages is a binary tree of the unit processing elements~\cite{gross_polar1, llr_list} (PEs) shown in Figs.~\ref{fig: psu_top}(b) and~\ref{fig: psu_top}(c), where $m$ is an integer. Stage $t$ ($t\geqslant m$) has $2^{n-t}$ PEs. Each of the remaining $m-1$ stages has the same circuit.

(b) Two types of PEs can be used in the PE tree in Fig.~\ref{fig: psu_top}(a). Suppose the maximal length of the constituent codeword that is decoded instantly or by the proposed LMLD algorithm in~\cite{jun_sips} is $2^\mu$, then stage $t$ ($t \geqslant n- \mu$) employs only type-I PEs. The other stages in the PE tree employ type-II PEs.

(c) Compared to the type-II PE, the type-I PE has an extra data load unit. For PE$_{l,t,j}$ within stage $t$, the binary outputs, $o_{l,t,2j}$ and $o_{l,t,2j+1}$, are connected to $b_{l,t-1, 2j}$ and $b_{l,t-1, 2j+1}$, respectively.

(d) BM$_{l,t}$ ($t \leqslant m-1$) is a bit memory with $\frac{2^{n-t}}{T}$ words, where each word contains $T$ bits. $T$ is the number of processing elements belonging to a decoding path in a partial parallel list decoder.

(e) The connector module (CN) has two $T$-bit inputs and two $T$-bit outputs. The connections between the outputs and inputs are given by
\begin{eqnarray}
\left\{\begin{array}{llll}
O_0[2j]& = & I_0[j] \oplus I_1[j] & 0\leqslant j < T/2\\
O_0[2j+1] & = & I_1[j] & 0\leqslant j < T/2 \\
O_1[2j-T] & = & I_0[j]  \oplus I_1[j] & T/2\leqslant j < T\\
O_1[2j+1-T] & = & I_1[j] & T/2\leqslant j < T
\end{array}
\right.\label{eqn: xor_eq}
\end{eqnarray}

(f) For each PE, $m_{l,t,j}$ in Figs.~\ref{fig: psu_top}(b) and~\ref{fig: psu_top}(c) is the output of an $L$-to-1 multiplexor whose inputs are $q_{0,t,j}$, $q_{1,t,j}$, $\cdots$, $q_{L-1,t,j}$. For each CN, $M_{l,t}$ is the output of an $L$-to-1 array whose inputs are $Q_{0,t}, Q_{1,t}, \cdots, Q_{L-1,t}$. These multiplexors are not shown in Fig.~\ref{fig: psu_top} for simplicity.


The proposed HPSU is derived from Alg.~\ref{algo: psum_comp}. For decoding path $l$, a round of partial sum computation is triggered once a constituent codeword $\mathcal{C}_{v,l}$ is decoded, where $\mathcal{C}_{v,l} = (c_{l,0}, c_{l,1},\cdots,c_{l,N_c-1})$ and $N_c=2^{n-t_v}$ is the length of the underlying constituent codeword. Suppose partial sums $(C_{l,t}[0][0]$, $C_{l,t}[1][0],$ $\cdots,$  $C_{l,t}[2^{n-t}-1][0])$ will be computed, where $t=I_e$ as shown in Alg.~\ref{algo: psum_comp}. The partial sum computation can be described as follows.

$\bullet$ For decoding path $l$, only $C_{l,t_v}, C_{l,t_v-1},\cdots,C_{l,t}$ are involved in the partial sum computation.

$\bullet$ For $l=0,1,\cdots, L-1$ and $k=n,n-1,\cdots,0$, let $\mathbf{C}_{l,k,0}$ and $\mathbf{C}_{l,k,1}$ denote two partial sum sets, where
     \begin{eqnarray}
     \begin{array}{l}
     \mathbf{C}_{l,k,0} = (C_{l,k}[0][0], C_{l,k}[1][0],\cdots,C_{l,k}[2^{n-k}][0]),\\ \nonumber
     \mathbf{C}_{l,k,1} = (C_{l,k}[0][1], C_{l,k}[1][1],\cdots,C_{l,k}[2^{n-k}][1]). \nonumber
     \end{array}
     \end{eqnarray}

$\bullet$ For $k=t_v-1$ to $t-1$, $\mathbf{C}_{l,k,1}$ is updated in serial during the partial sum computation. Here, $\mathbf{C}_{l,t_v,1}$ is initialized by the input constituent codeword $\mathcal{C}_{v,l}$, where $C_{l,t_v}[j][1] = c_{l,j}$ for $j=0,1,\cdots,2^{n-t_v}$.

$\bullet$ For $k=t_v$ to $t-1$, $\mathbf{C}_{l,k,0}$ remains unchanged during the current partial sum computation. However, $\mathbf{C}_{l,t,0}$ will be updated and used for the following LLR computation.

\begin{figure} [hbt]
\centering
  \includegraphics[width=2.2in]{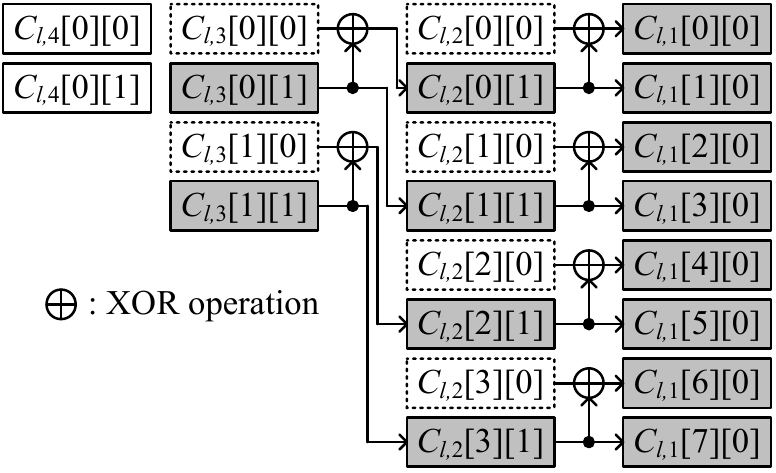}
  \caption{Schedule of partial sum computation when $n=4$, $t=1$ and $t_v=3$}\label{fig: psu_comp_flow}
\end{figure}

Let $n=4$, $t=1$ and $t_v=3$, the computation of partial sum sets $\mathbf{C}_{l,1,0}$ is shown in Fig.~\ref{fig: psu_comp_flow}, where the partial sums in shaded boxes will be updated and the partial sums in dash line boxes remain unchanged. Without loss of generality, we assume that the computation of $\mathbf{C}_{l,1,0}$ for decoding path $l$ is based on partial sums within path $l$ to simplify the discussion. The detailed computation is shown as follows.

$\bullet$ $\mathbf{C}_{l,3,1}$, which contains two partial sum bits, is first initialized with the input constituent codeword.

$\bullet$ $\mathbf{C}_{l,2,1}$ is computed based on the XOR network shown in Fig.~\ref{fig: psu_comp_flow}.

$\bullet$ The target partial sum set $\mathbf{C}_{l,1,0}$ is computed once $\mathbf{C}_{l,2,1}$ is updated.

For decoding path $l$, stage $t$ of the proposed PSCU stores only $\mathbf{C}_{l,t,0}$. When $t \geqslant m$, the single bit register D within PE$_{l,t,j}$ stores $C_{l,t}[j][0]$ for $j=0,1,\cdots,2^{n-t}$. When $t < m$, $(C_{l,t}[0][0]$, $C_{l,t}[1][0],$ $\cdots,$  $C_{l,t}[2^{n-t}-1][0])$ are stored in the bit memory BM$_{l,t}$, where the $k$-th word stores $(C_{l,t}[T(k-1)][0], C_{l,t}[T(k-1)+1][0],\cdots, C_{l,t}[T(k-1)+T-1][0])$.

For the proposed HPSU, the schedule of the computation of $\mathbf{C}_{l,t,0}$ depends on $t$. The detailed computation schedule is shown as follows.

(1) The decoded constituent codeword for decoding path $l$ is fed into the corresponding PSCU. Suppose the length of the constituent codeword is $N_c = 2^ {n-t_v}$. If the constituent codeword is from a rate-1 or ML node~\cite{jun_sips}, then LD$_{t_v}$ in Fig.~\ref{fig: psu_top}(b) is set to 0 to let the 2-to-1 multiplexor choose the constituent codeword input. Meanwhile, LZ$_{t_v}$ is set to 1. If the constituent codeword is from a rate-0 node~\cite{jun_sips}, LZ$_{t_v}$ is set to 0, since the corresponding constituent codeword is an all zero vector. LD$_t$ and LZ$_t$ for $t \neq t_v$ are both set to 1.

(2) When $t \geqslant m$, all $2^{n-t}$ partial sums belonging to $C_{l,t}$ are computed in one clock cycle. For stage $k$ with $t_v \geqslant k >t$, $m_{l,k,j}$ shown in Fig.~\ref{fig: psu_top}(b) and (c) is connected to $q_{p_l[k], k, j}$ due to the use of the lazy copy partial sum computation shown in Alg.~\ref{algo: psum_comp}, where $p_l[k]$ is a reference index. The partial sum output $s_{l,t,j}$ is just the updated $C_{l,t}[j][0]$ for $j=0,1,\cdots,2^{n-t}$.

(3) When $t<m$, the partial sums are generated in a partial-parallel way. Since there are only $T$ PUs for each decoding path, it needs at most $T$ partial sums per clock cycle~\cite{jun_low_mem_list, tree_list_dec}. Hence, at most $T$ partial sums are needed during each clock cycle.

\begin{figure} [hbt]
\centering
  \includegraphics[width=2.2in]{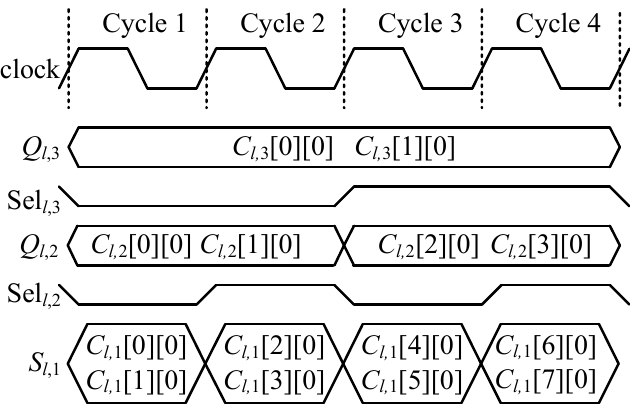}
  \caption{Partial parallel schedule of the partial sum computation example when $n=4$, $t=1$ and $t_v=3$}\label{fig: psu_mem_comp}
\end{figure}

Considering the partial sum computation shown in Fig.~\ref{fig: psu_comp_flow}, suppose $\mathbf{C}_{l,3,0}$, $\mathbf{C}_{l,2,0}$ and $\mathbf{C}_{l,1,0}$ are stored in bit memory BM$_{l,3}$, BM$_{l,2}$ and BM$_{l,1}$, respectively. Suppose $T=2$, the partial parallel computation of the $\mathbf{C}_{l,1,0}$ is shown in Fig.~\ref{fig: psu_mem_comp}. For $n=4$, $t=1$ and $t_v=3$, it takes $\frac{2^{4-1}}{2}=4$ clock cycles to compute all 8 partial sums within $\mathbf{C}_{l,1,0}$. For the PCSU architecture shown in Fig.~\ref{fig: psu_top}, suppose $S_{l,k}$, which has $T$ partial sums, is updated, the CN will generate 2$T$ partial sums.


Compared to the partial sum computation architectures in~\cite{tree_list_dec, jun_list}, the proposed HPSU architecture has advantages in the following two aspects.

(1) The proposed HPSU is a scalable architecture. The PSCU architectures in~\cite{tree_list_dec, jun_list} require $L(N-1)$ and $L(N/2-1)$ single bit registers, where $N=2^n$ is the block length. Hence, they will suffer from excessive area overhead when the block length $N$ is large. The proposed HPSU stores $L(N-1)$ bits and most of these bits are stored in RFs or SRAMs, which are more area efficient than registers. 

(2) The architectures in~\cite{tree_list_dec, jun_list} employ direct copying, which copies partial sums of a decoding path to another decoding path. In contrast, the proposed HPSU employs the lazy copy: it copies only index references. We define the copying of a single bit from one register to another as a single copy operation. Hence, when decoding path $l'$ needs to be copied to path $l$, the PSCU in~\cite{jun_list} requires $N_1 = 2^{n-1}-1$ copy operations, while the PSCU with lazy copy needs only $N_2 = (n+1)\log_2 L$ copy operations.
Since the value of $L$ for practical hardware implementation is small, our lazy copy needs much fewer copy operations than direct copy.

\section{Hardware Implementation results}\label{sec: hardware_imp}
In this paper, when $L=4$ and $T=128$, for $N=2^{13}$ and $2^{15}$, the proposed hybrid partial sum computation unit architecture is implemented with $m=3$ and $m=5$, respectively, under a TSMC 90nm CMOS technology. Our partial sum computation unit consumes an area of 0.779mm$^{2}$ and 1.31mm$^2$ for $N=2^{10}$ and $N=2^{15}$, respectively.

To the best of our knowledge, those decoder architectures in~\cite{tree_list_dec, jun_low_mem_list, chuan_list, yuan_low_latency} are the only for SC based list decoding algorithms of polar codes. However, in~\cite{tree_list_dec,chuan_list, yuan_low_latency}, the partial sum computation unit architecture was not discussed in detail and the implementation results on the PSCU alone are not shown. Hence, we compare our proposed PSCU with that in~\cite{jun_low_mem_list}.
When $L=4$, the partial sum unit architecture in~\cite{jun_low_mem_list} for $N=2^{13}$ and $2^{15}$  consumes an area of 1.011mm$^{2}$ and 3.63mm$^2$, respectively, under the same CMOS technology. All PSCUs are synthesized under a frequency of 500MHz. Our PSU achieves an area saving of 23\% and 63\% for block length $2^{13}$ and $2^{15}$, respectively.

For the list decoders in~\cite{jun_low_mem_list}, the area of the PSCU takes about 10\% of the overall decoder area for a polar code of block length $N=2^{10}$. This percentage will increase for a larger block length since the area of the register based PSCU increases more quickly than the rest of a list decoder. Thus, while our proposed PSCU will lead to area and energy saving for both long and short polar codes, the saving will be more significant for longer polar codes. Besides, the area saving also depends on $T$, since each bit memory could be implemented with an RF or a SRAM. As $T$ increases, the depth of a bit memory decreases. As a result, the area efficiency (total area normalized by total stored bits) decreases as shown in~\cite[Table I]{jun_low_mem_list}. The area saving does not depends on $L$.

By replacing the registers with memories, our PSCU does not introduce extra clock cycles for semi-parallel list decoder architectures~\cite{tree_list_dec, jun_low_mem_list} of polar codes. However, the critical path delay of our PSCU increases compared with that in~\cite{jun_list, tree_list_dec}.

\section{Conclusion}\label{sec: conclusion}

In this paper, a lazy copy partial sum computation algorithm is proposed. Based on this algorithm, a hybrid partial sum computation unit architecture is also proposed. Compared with existing architectures, our architecture is more area efficient and energy efficient by eliminating the copy of partial sums.

\bibliographystyle{IEEEbib}
\bibliography{refs_latest}

\end{document}